\documentclass[twocolumn,showpacs,preprintnumbers,amsmath,amssymb,prl,superscriptaddress]{revtex4-1}
\pdfoutput=1
\usepackage{graphicx}
\usepackage{dcolumn}
\usepackage{bbold}
\usepackage{color}

\usepackage[low-sup]{subdepth}
\usepackage{multirow}
\usepackage{dsfont}
\usepackage{amsfonts}
\usepackage[english]{babel}
\usepackage[T1]{fontenc}
\usepackage{times}
\usepackage[colorlinks,bookmarks=true,citecolor=blue,linkcolor=red,urlcolor=blue]{hyperref}
\usepackage[tight, FIGTOPCAP, hang, raggedright, nooneline]{subfigure}

\usepackage{verbatim}
\usepackage{bm}
\usepackage{mathrsfs}

\newcommand* {\co}[1]{\hat{c}_{#1}}
\newcommand* {\codag}[1]{\hat{c}^\dag_{#1}}
\newcommand* {\n}[1]{(\hat{n}_{#1}-1)}

\usepackage{amsmath,amssymb}
\usepackage{color}
\usepackage{graphicx}
\usepackage{natbib}

\begin{document}
\title{Hubbard Model on the Honeycomb Lattice with an Indefinite Long-Range Interaction}
\author{Mohammad-Sadegh Vaezi}
\email{ms.vaezi@khatam.ac.ir}
\email{sadegh.vaezi@gmail.com}
\affiliation{Department of Converging Technologies, Khatam University, Tehran 19395-5531, Iran}
\affiliation{Pasargad Institute for Advanced  Innovative Solutions (PIAIS), Tehran 19395-5531, Iran}
\author{Davoud Nasr Esfahani}
\email{d.nasr@khatam.ac.ir}
\email{dd.nasr@gmail.com}
\affiliation{Department of Converging Technologies, Khatam University, Tehran 19395-5531, Iran}
\affiliation{Pasargad Institute for Advanced  Innovative Solutions (PIAIS), Tehran 19395-5531, Iran}


\begin{abstract}
Several studies have emphasized the impact of long-range Coulomb interactions in lattice fermions, yet conventional Auxiliary Field Quantum Monte Carlo (QMC) methods face limitations due to their reliance on positive definite interaction matrices.
We address this by decomposing the interaction matrix into positive- and negative-definite components, allowing for QMC calculations with manageable sign properties. This technique enables effective simulations on large lattices. Applying it to a honeycomb lattice with an indefinite interaction matrix, we identify a semi-metal to charge density wave phase transition within the Gross-Neveu criticality class. 
Notably, the phase transition boundary aligns with regions where the average sign sharply decreases, providing new evidence to the increasingly compelling research on the relationship between phase transitions and the sign problem.
\end{abstract}

\maketitle


\textit{\textbf{Introduction.}}\,---\, Two-dimensional (2D) Dirac fermions have been a subject of investigation since their initial theoretical predictions~\cite{semenoff1984condensed,gross1974dynamical}. The discovery and subsequent experimental realization of them in materials like graphene have spurred extensive research efforts to explore their unique properties and interaction-driven phenomena. These systems, characterized by a linear dispersion relation near the Dirac points, exhibit remarkable electronic behavior, making them a focal point in condensed matter physics~\cite{RevModPhys.84.1067,schuler2013optimal,elias2011dirac,faugeras2015landau,ulybyshev2013monte,wehling2014dirac,tang2018role,he2018dynamical,PhysRevLett.123.137602,tang2024spectral}.

\indent A key area of interest has been understanding the stability of the gapless Dirac state under various interaction conditions. It has been established that while the gapless state is stable against weak interactions, a phase transition to a gapped state can occur at finite interaction strengths~\cite{herbut2006interactions}. This transition has been extensively studied using Quantum Monte Carlo (QMC) simulations, particularly in the context of the Hubbard and spinless fermion models on a honeycomb lattice~\cite{assaad2013pinning,sorella2012absence,costa2021magnetism,sorella1992semi,toldin2015fermionic,otsuka2020dirac,ostmeyer2020semimetal,capponi2016phase,wang2014fermionic,raczkowski2020hubbard}, where the local interaction $U$ induces a quantum phase transition from a semi-metal (SM) phase to a gapped antiferromagnetic (AFM) spin density wave (SDW) phase, with critical exponents consistent with the Gross-Neveu universality class~\cite{PhysRevLett.128.225701,boyack2021quantum,PhysRevLett.123.137602,PhysRevLett.128.087201}.
The impact of long-range Coulomb interactions (LRC) on these systems has also been a major focus. Renormalization Group (RG) analysis and QMC simulations have shown that while LRC interactions shift $U_c$ to larger values, they do not change the universality class of the transition~\cite{tang2018role,herbut2006interactions,de2017competing,hohenadler2014phase,schuler2013optimal}.
Mean-field approaches~\cite{araki2012spin} corroborate these findings.
Further studies suggest that in the phase space of the local and non-local parts of the interaction, a transition from a SM to a charge density wave (CDW) phase can occur \cite{de2017competing,classen2015mott,classen2016competition} (akin to  Hubbard-Holstein model~\cite{chen2019charge,xiao2019competition,zhang2019charge,costa2020phase,dos2023phase,xiao2019competition,cohen2022fast,berger1995two,wang2015phonon,nowadnick2012competition,ohgoe2017competition,dos2023phase}).

\indent Despite these significant advances, accurately modeling Coulomb interaction-driven transitions, particularly in Auxiliary Field QMC (AFQMC) simulations, remains a challenging endeavor. In the discrete version of AFQMC, the number of auxiliary fields increases rapidly with the number of interactions, leading to a substantial rise in computational complexity and operational time. Moreover, the sign problem (SP) intensifies in this context, restricting their applicability. Consequently, studies have often been restricted to considering only nearest-neighbor interaction terms~\cite{wu2014phase,chang2024boosting}.
In contrast, the continuum version of AFQMC allows for the inclusion of interactions beyond the nearest neighbors, provided the interaction matrix remains positive-definite. However, when the interaction is not positive-definite, i.e., indefinite, the method fails~\cite{tang2018role}. 
Decomposing the interaction matrix into positive- and negative-definite components by adding and subtracting a diagonal matrix might seem straightforward~\cite{sugiyama1986auxiliary}. However, as we will see later, this approach suffers from a severe SP.

\indent In this manuscript, we propose a new decomposition method based on the eigenvalues of the interaction matrix, which provides a simple yet effective fix for the issues associated with an indefinite interaction matrix. Although our method introduces SP, it significantly reduces its severity, making it manageable for a wide range of lattice sizes and temperatures. This advancement enables us to determine the phase transition between the SM and CDW phases. We have implemented this method within the ALF (Algorithms for Lattice Fermions) framework \cite{bercx2017alf}, demonstrating its effectiveness in practical QMC simulations.
We should note that this method is generic and not restricted to the honeycomb lattice (see our results for the square lattice in the Supplemental Material~\cite{SuppMat}).

\indent Furthermore, we explore the correlation between the severity of the SP and the phase boundaries of the model, finding that our results are consistent with correlations observed in other models \cite{mondaini2022quantum,PhysRevB.107.245144,PhysRevB.106.L241109}.

~\\
\textit{\textbf{Model $\&$ Method.}}\,---\, We investigate the LRC Hubbard model on a honeycomb lattice with $N = 2 L_1 \times L_2$ sites using the continuum version of the AFQMC method. Here, $L_1$ and $L_2$ represent the lattice dimensions along the basis vectors $a_1 = (1, 0)$ and $a_2 = \frac{1}{2}(1, \sqrt{3})$, respectively.
The Hamiltonian is given by
\begin{eqnarray}
\hat{H} = -t\sum_{\langle ij\rangle,\sigma} \codag{i\sigma}\co{j\sigma} + \text{h.c.} + \frac{1}{2}\sum_{ij}V_{ij}\n{i}\n{j},
\end{eqnarray}
where $i =({\bf i, \delta_i})$ and $j =({\bf j, \delta_j})$ are super-indices that contain both the unit cell and orbital information. Furthermore, $\langle ij\rangle$ denotes nearest-neighbor pairs, $\co{i\sigma}$ is the annihilation operator for a spin-$\sigma$ electron at site $i$, $\hat{n}_i = \hat{n}_{i\uparrow} + \hat{n}_{i\downarrow} $, and $\hat{n}_{i\sigma} = \codag{i\sigma}\co{i\sigma}$ is the number operator.
We assume $t=1$ throughout this paper.
The interaction is defined as~\cite{bercx2017alf}
\begin{eqnarray}
V_{i,j} \equiv V(\bf{i+\delta_i, j+\delta_j}) = U \times \left\{
\begin{array}{ll}
1 & ~ i = j, \\
 \frac{\alpha d_{\text{min}}}{\|\bf{i - j + \delta_i - \delta_j}\|} & ~ i\neq j,
\end{array}
\right.
\end{eqnarray}
where $d_{\text{min}} = 1/\sqrt{3}$ denotes the minimum distance between two orbitals. We note that, with periodic boundary conditions (as assumed in this study), $\|r\|$ represents the minimum distance between two points on the torus~\cite{bercx2017alf}. 
Here, $\alpha$ controls the strength of long-range interactions. Notably, the $V$ matrix remains positive-definite for $\alpha < \alpha_c$ independent of $U$. In the Supplemental Material~\cite{SuppMat}, we analyze $\alpha_c$ for various lattice sizes and find very small variation.
To ensure consistency, we will set $\alpha_c = 0.6484$ throughout the paper. This choice poses no issues, as applying our method to a positive-definite $V$ does not induce an SP (for the parameters studied in this work).

\indent This model has been exclusively analyzed using AFQMC when $V$ is positive-definite~\cite{tang2018role,sorella2012absence}. Here, we extend this approach to indefinite cases.
To achieve this, one must first decompose the interaction matrix as $V = V^{+} + V^{-}$, where $V^{+}$ and $V^{-}$ are purely positive- and negative-definite matrices, respectively, that need to be determined.
This allows us, when computing the partition function, to apply the Hubbard-Stratonovich transformation to each component separately—one with imaginary auxiliary fields and one with real fields~\cite{SuppMat}—to avoid divergence. That is,
\begin{eqnarray}
&~&e^{\frac{-\Delta \tau}{2} \hat{n}_{i}{V_{ij}^{+}}\hat{n}_{j}} \nonumber \\
&\propto& \int \prod_{i} d \phi^+_{i}   e^{ - \frac{\Delta \tau}{2} \sum_{i,j} \phi^+_{i} {V_{ij}^{+}}^{-1}  \phi^+_{j} - \sum_{i}  i \Delta \tau \phi^+_i (\hat{n}_{i}-1) }, \nonumber \\
&~&e^{\frac{-\Delta \tau}{2} \hat{n}_{i}{V_{ij}^{-}}\hat{n}_{j}} \nonumber \\
&\propto& \int \prod_{i} d \phi^-_{i}   e^{\frac{\Delta \tau}{2} \sum_{i,j} \phi^-_{i} {V_{ij}^{-}}^{-1}  \phi^-_{j} - \sum_{i}   \Delta \tau \phi^-_i (\hat{n}_{i}-1) }.
\end{eqnarray}
Here, the time step $\Delta \tau = \beta / L_{\text{Trotter}}$, the inverse temperature $\beta = 1 / k_B T$, and $L_{\text{Trotter}}$ denotes the number of time steps (in the numerical result we set $\Delta \tau = 0.1$ giving rise to a good data convergence).
Please note that $\phi^{+}$ and $\phi^{-}$ are two distinct sets of auxiliary fields (for more information regarding the behavior of these fields and other details please see Ref.~\cite{bercx2017alf}).
Now, using the second order symmetric Trotter decomposition, the partition function can be expressed as
\begin{eqnarray} \label{eqn:LRCPatitionFun}
Z &\propto& \int \prod_{i,\tau,\nu=\pm} d \phi^{\nu}_{i, \tau} e^{ - \frac{ \nu \Delta \tau} {2} \sum_{i,j} \phi^{\nu}_{i,\tau} {V_{ij}^{\nu}}^{-1}\phi^{\nu}_{j,\tau}} \nonumber \\
&\times& \text{Tr} \Big[ \prod_{\tau}
e^{-\frac{\Delta \tau \hat{H}_T}{2}}  e^{- \sum_{i} \Delta \tau
(\imath\phi^{+}_{i,\tau}+\phi^{-}_{i,\tau}) (\hat{n}_{i}-1)  } e^{-\frac{\Delta \tau \hat{H}_T}{2}} \Big],\nonumber\\
\end{eqnarray}
where $\hat{H}_T = -t\sum_{\langle ij\rangle,\sigma} \codag{i\sigma}\co{j\sigma}$ is the the hopping Hamiltonian. The remaining task is to determine $V^{+}$ and $V^{-}$. Here, we discuss two methods for performing this decomposition:

{\bf 1.} "Direct Decomposition" (DD): We add and subtract a constant $\epsilon$ from the $V$ matrix so that $V^{+} = \epsilon\mathds{1} + V$ and $V^{-} = -\epsilon\mathds{1}$~\cite{SuppMat, sugiyama1986auxiliary}. Here, $\epsilon$ is some real positive number. 

{\bf 2.} "Eigenvalue Decomposition" (ED): We use the eigenvalues of the $V$ matrix for the decomposition. Suppose the $V$ matrix has $M$ negative eigenvalues out of $N$. Then, $V^{+} = W (\epsilon\mathds{1} + \Lambda^{+}) W^T$ and $V^{-} = W (-\epsilon\mathds{1} + \Lambda^{-}) W^T$, where $W$ is an orthonormal matrix that diagonalizes $V$. The diagonal matrices $\Lambda^{+}$ and $\Lambda^{-}$ are defined as:
\begin{eqnarray}
\Lambda^{-} &=& \text{Diag}\left( \lambda_1, \cdots, \lambda_{M}, 0, \cdots, 0 \right), \nonumber \\
\Lambda^{+} &=& \text{Diag}\left( 0, \cdots, 0, \lambda_{M+1}, \cdots, \lambda_{N}\right),
\end{eqnarray}
with $\lambda_1, \cdots, \lambda_{M} < 0$ and $\lambda_{M+1}, \cdots, \lambda_{N} > 0$, and $\epsilon$ is some real positive number. 

\indent Notably, for the DD method, there exists a finite lower bound, $\epsilon_{\text{min}}$, such that $V^{+}$ and $V^{-}$ are only defined for $\epsilon > \epsilon_{\text{min}}$, where $\epsilon_{\text{min}}$ depends on the structure of $V$~\cite{SuppMat}. In contrast, for the ED method, $\epsilon$ can be any positive real number.
We will demonstrate (see numerical results) that the ability to freely select $\epsilon$ in the ED method leads to a significant improvement in the SP compared to the DD method.
We would like to point out that Ref.~\cite{hamann1993long} employs an eigenvalue-based decomposition technique within the hybrid QMC method. However, there are notable differences between their approach and ours, including the number of auxiliary fields per site, the specific method utilized, and various technical aspects. Most importantly, our method uses fewer auxiliary fields, which leads to reduced computational time and complexity, ultimately enhancing overall efficiency.

~\\
\textit{\textbf{Monte Carlo Sampling.}}\,---\, 
Following Ref.~\cite{hohenadler2014phase}, we use sequential single spin flips combined with global spatial updates. Below, we will briefly describe the Monte Carlo sampling process for our modified partition function. For the updates in our LRC Hubbard model, whose partition function is given in Eq.~\ref{eqn:LRCPatitionFun}, it is convenient to work in the basis where $V$ is diagonal. Thus, we define $\gamma^{\pm}_{i,\tau} = \sum_{j} W^{T}_{i,j} \phi^{\pm}_{j,\tau}$.
On a given time slice $\tau_u$ we propose a  new field configuration with the probability 
$T^{0}_{\Lambda  \rightarrow  \Lambda'} =
\prod_{i} \left[  p P_B({\gamma'}^\nu_{i,\tau_u})  + (1-p) \delta( \gamma^\nu_{i,\tau_u} - {\gamma'}^\nu_{i,\tau_u})   \right]$ for $\tau = \tau_u$, and
 $T^{0}_{\Lambda  \rightarrow  \Lambda'} = \delta( \gamma^\nu_{i,\tau} - {\gamma'}^\nu_{i,\tau})$ for
  $\tau \neq \tau_u$, where $P_B(\gamma^{\nu}_{i,\tau})  \propto e^{\frac{-\Delta \tau}{2|\Lambda^{\nu}_i|} {{\gamma^{\nu}_{i,\tau}}^2}}$.
Here, $\nu = \pm$, $p \in [0,1]$, and $\delta$ denotes the Dirac $\delta$-function. In other words, we randomly sample the field with probability $p$ and keep the field unchanged with probability $(1 - p)$. The parameter $p$ is adjustable to control the acceptance rate but should not affect the final results. In our analyses, setting $p = 0.3$ achieved satisfactory convergence within a reasonable computational time. 
Additionally, we employ the Metropolis-Hastings acceptance-rejection scheme and accept the move with probability
$\text{min} \left(\frac{T^{0} ( \Gamma'  \rightarrow \Gamma ) W_B(\Gamma') W_F(\Gamma') }{ T^{0} ( \Gamma \rightarrow \Gamma' ) W_B(\Gamma) W_F(\Gamma) }, 1 \right) =
\text{min} \left(\frac{W_F(\Gamma') }{ W_F(\Gamma) }, 1 \right)$,
where 
$W_B(\Gamma) = e^{-\Delta \tau \sum_{i,\tau,\nu=\pm} \frac{{\gamma^{\nu}_{i,\tau}}^2}{2|\Lambda^{\nu}_i|}}$, and
$W_F(\Gamma) = \text{Tr} \left[   \prod_{\tau}
e^{-\Delta \tau \hat{H}_T}  e^{- \sum_{i,j}  \Delta \tau W_{i,j} (\gamma^{-}_{j,\tau} + i \gamma^{+}_{j,\tau}) \left( \hat{n}_{i} - 1 \right) }\right]$.
It is important to note that for a given time slice, a local modification in the diagonal basis ($\gamma^{\pm}$) corresponds to a non-local spatial update in the non-diagonal basis ($\phi^{\pm}$).

\begin{figure}
\includegraphics[scale=0.34]{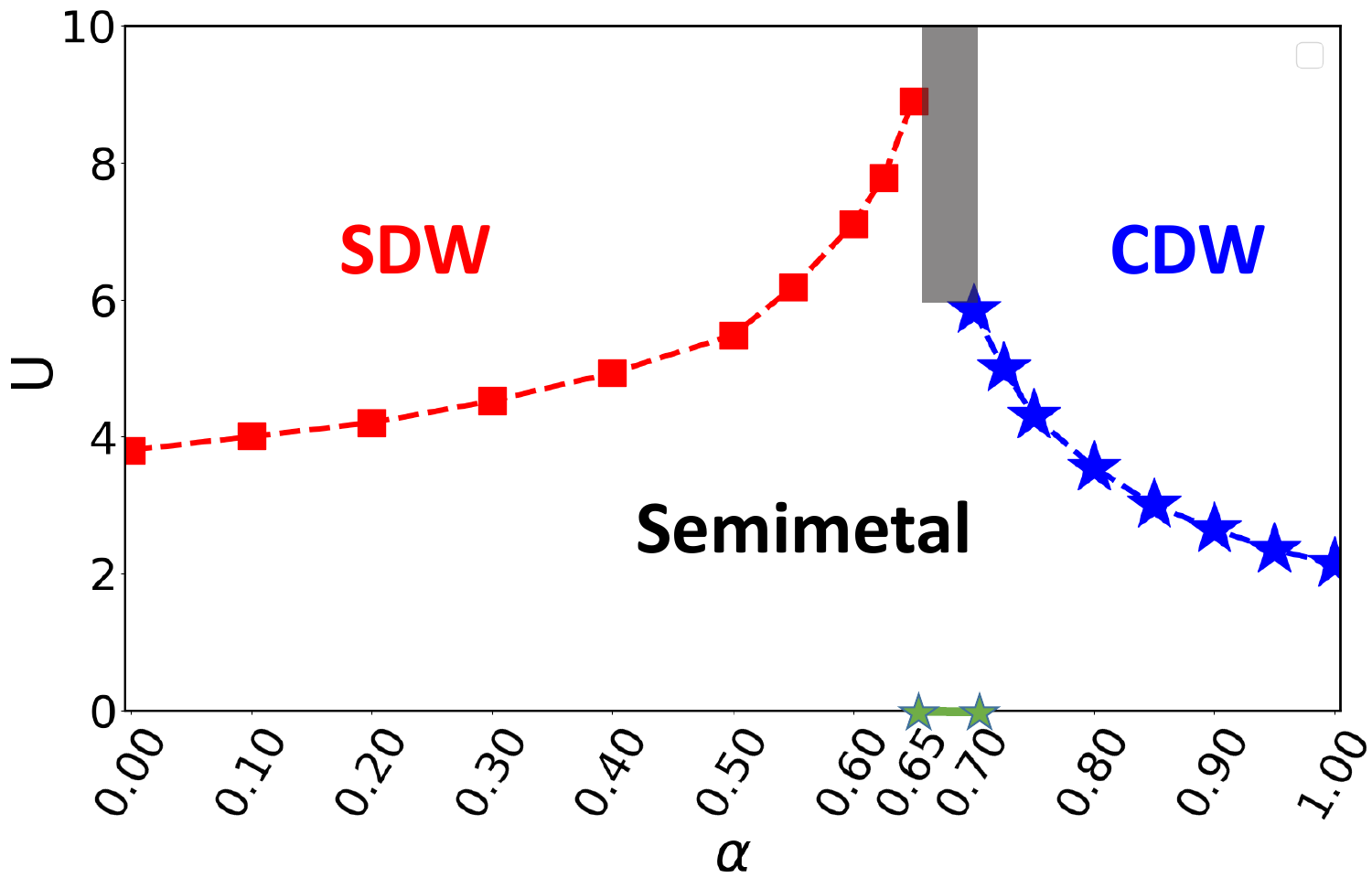}
\caption{The phase diagram of the honeycomb lattice as a function of the local interaction parameter $U$ and the long-range Coulomb interaction parameter $\alpha$. The shaded areas indicate regions where the sign problem is particularly severe, rendering the data unreliable and challenging to interpret. For details, see the main text.}
\label{phasediag}
\end{figure}
\begin{figure}
\includegraphics[scale=0.5 ]{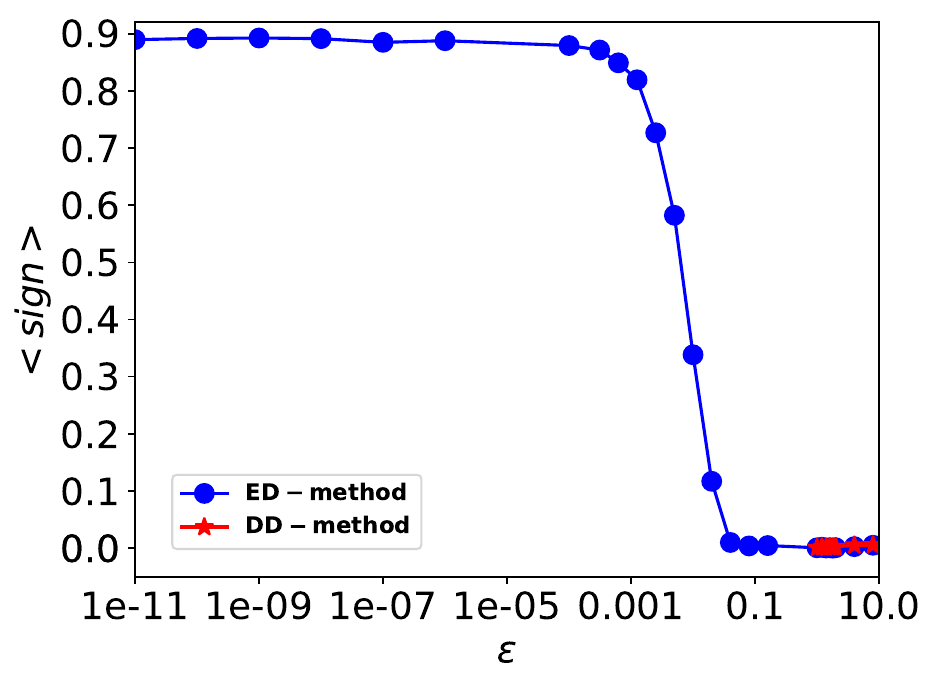}
\caption{Comparison of the average sign for the Direct and Eigenvalue based decomposition of interaction matrix, for a lattice with $\alpha=0.8$, $U=3.4$, $L=9$, and $\beta=12.0$.}
\label{sign}
\end{figure}
\begin{figure}
\includegraphics[scale=0.4]{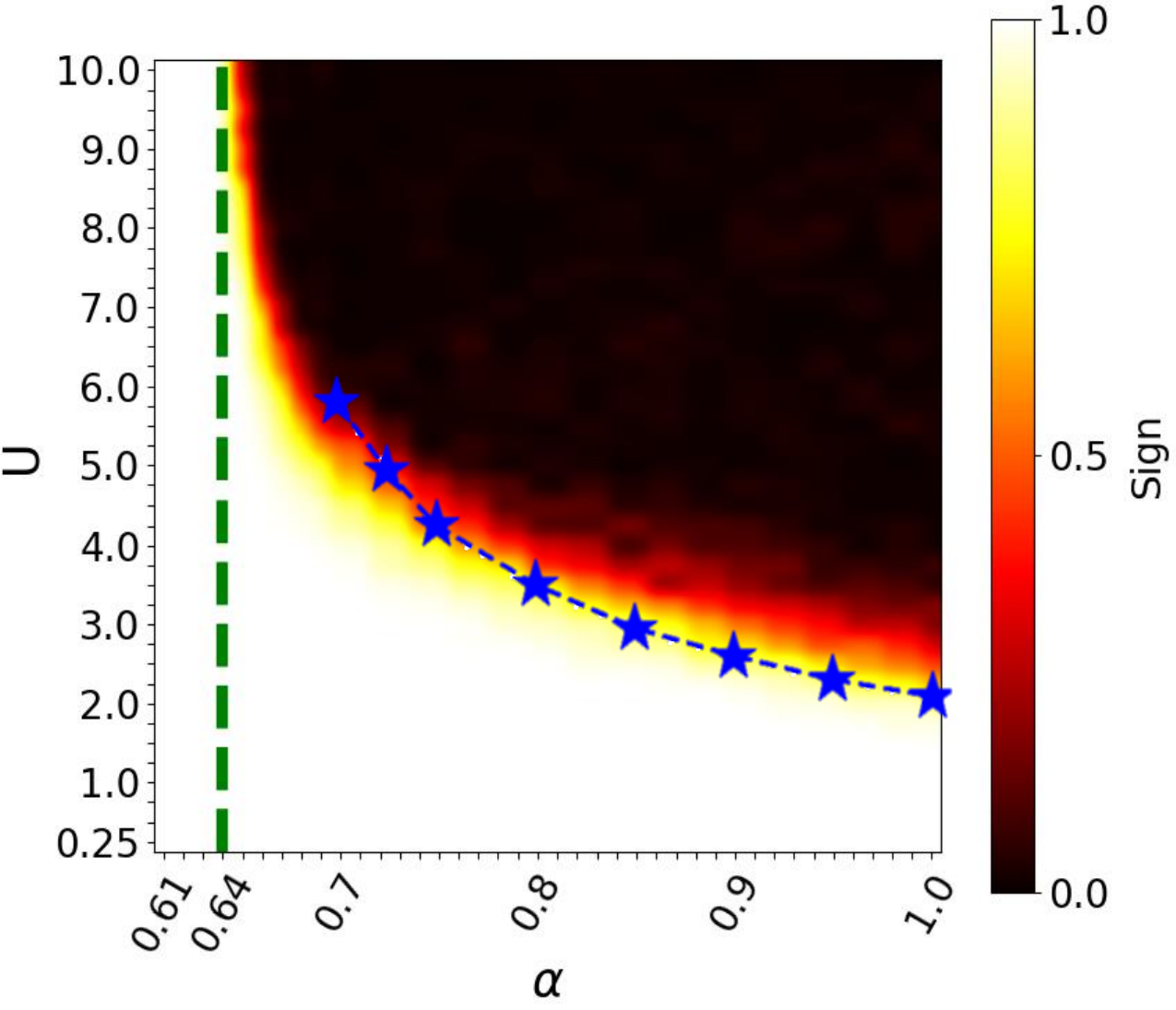}
\caption{The contour plot illustrates the sign behavior for $\beta=12$, $L=12$ concerning $U$ and $\alpha$. A distinct red region is noticeable, delineating a region where the sign approximately equals 1 from another where it approaches 0. This pattern qualitatively resembles the boundary observed between SM and CDW phases, depicted in Fig.~\ref{phasediag}.}
\label{heatmap}
\end{figure}
\begin{figure*}[!t]
\includegraphics[width=0.85\textwidth]{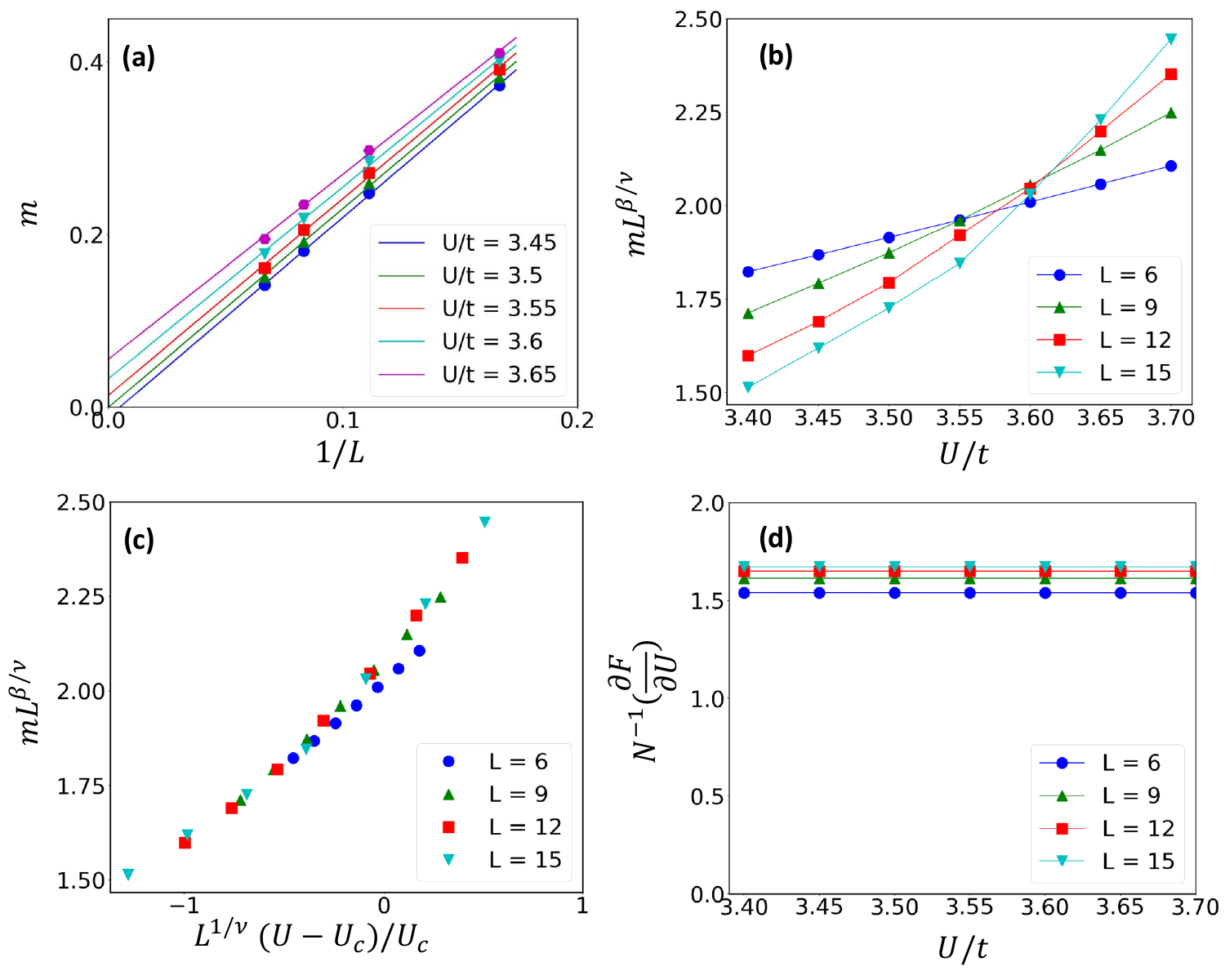}
\caption{
(a) Finite size scaling of $m_{CDW}$ for $\alpha = 0.8$ and different values of $U$.
(b) Plotting $mL^{\beta/\nu}$ against $U$, where $\beta/\nu = 0.9$ and $\nu= 0.88$, consistent with the critical exponents of the Gross-Neveu criticality class (please note that here the $\beta$ refers to the critical exponents and not the inverse temperature). The curves corresponding to various sizes intersect at $U_c \approx 3.615$.
(c) The convergence of curves onto a single curve for the parameters provided in panel (b) confirms the Gross-Neveu universality class.
(d) The smooth variation of $N^{-1}{\frac{\partial F}{\partial U}}$ with respect to both $U$ and $L$ provides compelling evidence for a second-order phase transition.}
\label{grossnova}
\end{figure*}
~\\
\textit{\textbf{Numerical Results.}}\,---\, To investigate different phases of this model, we utilize the structure factor
\begin{eqnarray}
S_{X}(k) &=& \frac{1}{N}\sum_{ij} e^{i k.(i-j)} \nonumber\\
&\times&\langle (O^X_{A,i}-O^X_{B,i})(O^X_{A,j}-O^X_{B,j}) \rangle,
\end{eqnarray}
where $X=\text{CDW,~SDW}$, with $O_{\alpha,i}^{\text{SDW}} = S^{z}_{\alpha,i} = \frac{1}{2}(\hat{n}_{\alpha,i\uparrow}-\hat{n}_{\alpha,i\downarrow})$ and $O_{\alpha,i}^{\text{CDW}} = \hat{n}_{\alpha,i\uparrow} + \hat{n}_{\alpha,i\downarrow}$. Here, $i$ and $j$ represent unit-cell indices, while $\alpha = \text{A,B}$ denotes the sub-lattice indices within each unit cell. For the SDW and CDW, which are expected to manifest at the $\Gamma$ point, the order parameters considered throughout this manuscript are defined as
\begin{eqnarray}
   m_{X} = \sqrt{\frac{S_{X}(\Gamma)}{N}}.
\end{eqnarray}
\indent As mentioned earlier, the phase diagram of this model is well-established when $V$ is positive-definite, exhibiting a transition from SM to SDW phase~\cite{tang2018role,sorella2012absence}. That is, $m_{\text{CDW}} = 0$ throughout, while $m_{\text{SDW}}$ takes zero and nonzero values before and after the transition, respectively.
For completeness, we have regenerated the phase diagram (for a broader range of parameters compared to previous studies) as shown in Fig.~\ref{phasediag} for $0 \leq \alpha \leq \alpha_c$. Since the model is free of SP in this regime, we employed the Projective AFQMC method with $2\Theta + \beta = 40$ (see the Supplemental Material~\cite{SuppMat} and Ref.~\cite{bercx2017alf} for details) which better captures the zero temperature behavior of the system. A well-known observation is that incorporating the long-range Coulomb interaction (i.e., increasing $\alpha$) leads to a higher critical onsite interaction $U_c$.

\indent For cases where $\alpha > \alpha_c$, rendering $V$ indefinite, either the DD or ED method is required. To show that the ED method generally outperforms the DD method, we consider a specific parameter set: $\alpha = 0.8$, $U = 3.4$, $L = 9$, and $\beta = 12$. A comparison of the average sign (⟨sign⟩) between these methods is shown in Fig.~\ref{sign}.
In the DD method, as expected, the minimum threshold $\epsilon_{min} = 1.0$ imposes a lower bound, below which a $V^+,~V^-$ decomposition is not feasible. Setting $\epsilon \geq \epsilon_{min}$ leads to severe SP, effectively preventing meaningful simulations of the model.
In contrast, with the ED method, $\epsilon_{min}$ can be chosen arbitrarily small, which significantly mitigates the SP as $\epsilon \rightarrow 0$. Evidently, ⟨sign⟩ converges as $\epsilon$ is reduced below a certain point~\cite{SuppMat}. For large $\epsilon$ values, however, the ED method behaves similarly to the DD method.
Overall, we found that, across various parameters (e.g., different $\beta$ values, lattice sizes, and etc), setting $\epsilon = 10^{-10}$ achieves a good balance; it provides stable data convergence, maintains a high ⟨sign⟩, and avoids numerical instability from machine precision limitations.

\indent This method allowed us to identify the SM-CDW transition for $\alpha > \alpha_c$, as shown in Fig.\ref{phasediag}, with a well-defined $U_c(\alpha)$, marked by blue stars. 
It is important to mention that in the SM region, away from the SM-CDW boundary, the SP is minimal, allowing us to use the Projective AFQMC method. However, near the critical points, the SP intensified, forcing us to employ finite but sufficiently low-temperature to determine the phase boundary. 
Despite these efforts, the SP remains severe near the SDW-SM-CDW critical point, rendering calculations in this region impractical, as highlighted by the gray rectangle in Fig.~\ref{phasediag}.
A schematic illustration of the sign behavior is provided in Fig.~\ref{heatmap}.
Integrating SP amelioration techniques like Adiabatic QMC~\cite{vaezi2021amelioration} may improve exploration of these regions (will be addressed in future work).
Overall, in contrast to the case of $\alpha \leq \alpha_c$, increasing $\alpha$ in this regime results in a lower critical onsite interaction, $U_c$.

\indent To better illustrate our findings, we focus here, for instance, on the case of $\alpha = 0.8$, using the lowest temperature we could achieve in the finite temperature AFQMC method; $k_B T = 1/\beta = 1/12$. We evaluated $m_{\text{CDW}}$ for lattice sizes $L_1 = L_2 = L = 6, 9, 12, 15$ and $U$ values ranging from 3.45 to 3.64. 
The results, extrapolated to $L = \infty$ as a function of $\frac{1}{L}$ (see Fig.~\ref{grossnova}(a)), show a finite $m_{\text{CDW}}$ beyond a critical $U$ that clearly signals the SM-CDW transition.
It is worth noting that here the $m_{\text{SDW}}\sim 0$ for different values of $U$.
To accurately determine the critical point, we plotted $mL^{\beta/\nu}$ (where $m \equiv m_{\text{CDW}}$) versus $U$ in Fig.\ref{grossnova}(b), using critical exponents $\beta/\nu = 0.9$ (please note that this $\beta$ refers to the critical exponents and not the inverse temperature) and $\nu = 0.88$, consistent with Gross-Neveu criticality. The intersection of curves for different sizes at $U_c \approx 3.615$ identifies the transition. Further, we plotted $mL^{\beta/\nu}$ as a function of $L^{1/\nu}\times(U-U_c)/U_c$ in Fig.\ref{grossnova}(c), with the resulting data collapse confirming the Gross-Neveu phase transition.
Lastly, the second-order nature of this transition is supported by Fig.~\ref{grossnova}(d), where $N^{-1}{\frac{\partial F}{\partial U}}$ is plotted against $U$ for various sizes. The smooth variation of $N^{-1}{\frac{\partial F}{\partial U}}$ with $U$ and $L$ reinforces the second-order character of the phase transition.

\indent Finally, we highlight the recent observation of a correlation between the SP and phase transitions, a topic of considerable interest. In our model, we explored this correlation by examining the SP characteristics. Specifically, for $L=12$ and $\beta=12$, $\epsilon = 10^{-10}$, we present a contour plot of ⟨sign⟩ as a function of $U$ and $\alpha$ in Fig.~\ref{heatmap}. The SM-CDW phase transition boundary, marked by blue stars on the plot, generally aligns with regions where ⟨sign⟩ sharply decreases as we move from the SM phase. While it might seem that the choice of the non-physical parameter $\epsilon$ could disrupt this alignment, it’s important to note that, as shown in Fig.~\ref{sign} and in ~\cite{SuppMat}, ⟨sign⟩ converges below a certain threshold of $\epsilon$, beyond which it remains stable. Our choice of $\epsilon = 10^{-10}$ was made to ensure we are well below this threshold, as detailed further in the Supplemental Material~\cite{SuppMat}.
Additionally, the region where the ⟨sign⟩ is suppressed narrows at points where transitions occur at smaller $\alpha$ and larger $U$. This pronounced drop in ⟨sign⟩ makes it challenging to accurately trace the SM-CDW phase transition within the framework of Gross-Neveu criticality. We observed similar results for lower (larger) values of $\beta$ and $L$, though with a smoother (sharper) boundary~\cite{SuppMat}. 
The green dashed line is conjectured to mark the boundary between the SDW and CDW phases for large values of $U$.

~\\
\textit{\textbf{Conclusion.}}\,---\, We investigated the effect of long-range Coulomb interactions in the Hubbard model on a honeycomb lattice at half-filling using the AFQMC method. We tackled the challenge posed by the indefinite nature of the interaction matrix, which traditionally complicates such simulations. By decomposing the interaction matrix into positive- and negative-definite components, we were able to perform practical AFQMC simulations in previously inaccessible regimes.
Applying this method, we identified a SM-CDW phase transition within the Gross-Neveu criticality class when non-local interactions are dominant.
In contrast to the positive-definite region, incorporating long-range Coulomb interaction here lowers the critical onsite interaction, $U_c$.

\indent Additionally, we investigated the relationship between the severity of the sign problem and phase transitions. Our results reveal that the phase transition boundary generally aligns with regions where the sign problem experiences a significant reduction. This insight deepens our understanding of the interplay between the sign problem and phase transitions in complex quantum many-body systems.


~\\
\textit{\textbf{Acknowledgment.}}\,---\,We are grateful to Abolhassan Vaezi for valuable discussions throughout the course of this research.


\nocite{apsrev41Control}
\bibliographystyle{apsrev4-1}

%


\clearpage
\onecolumngrid
\appendix
\begin{center}
\textbf{\large Supplemental Material for "Hubbard Model on the Honeycomb Lattice with an Indefinite Long-Range Interaction"}
\end{center}
\setcounter{section}{0}
\setcounter{equation}{0}
\setcounter{figure}{0}
\setcounter{table}{0}
\setcounter{page}{1}
\makeatletter

\renewcommand{\thetable}{S\arabic{section}}
\renewcommand{\thetable}{S\arabic{table}}
\renewcommand{\theequation}{S\arabic{equation}}
\renewcommand{\thefigure}{S\arabic{figure}}

\subsection{ Mohammad-Sadegh Vaezi,~~Davoud Nasr Esfahani}
~~\\

\subsection{1) Hubbard-Stratonovich Transformation: Handling positive- and negative-definite Cases}

\indent To understand better our method for dealing with indefinite interaction matrices, it's essential to review some fundamental mathematical principles.
We begin by considering two familiar Gaussian integrals:
\begin{align}
\text{\bf (I)} &\quad \int e^{-a\phi_{i}^2 - i\phi_{i}y_{i}} \, d\phi_{i} 
&&\quad a>0,~\text{Gaussian integral for imaginary fields,}\nonumber \\
\text{\bf (II)} &\quad \int e^{a\phi_{i}^2 - \phi_{i}y_{i}} \, d\phi_{i} 
&&\quad a<0,~\text{Gaussian integral for real fields}.
\end{align}
Rewriting these integrals in a squared form and noting that $
\int e^{-|a|(x+b)^2} dx = \sqrt{\frac{\pi}{|a|}}$, we end up with:
\begin{eqnarray} \label{eqn: Gaussian imaginary integral revised a}
a>0 : &~& \int e^{-a\phi_{i}^2 - \mathbf{i}\phi_{i}y_{i}} \, d\phi_{i} = \int e^{-a\left(\phi_{i} + \frac{\mathbf{i}y_i}{2a}\right)^2 - \frac{y_i^2}{4a}} d\phi_{i} = \sqrt{\frac{\pi}{a}} e^{-\frac{y_i^2}{4a}}, \nonumber\\
a<0 : &~& \int e^{a\phi_{i}^2 - \phi_{i}y_{i}} \, d\phi_{i} = \int e^{a\left(\phi_{i} - \frac{y_i}{2a}\right)^2 - \frac{y_i^2}{4a}} \, d\phi_{i} = \sqrt{\frac{\pi}{-a}} e^{-\frac{y_i^2}{4a}}.
\end{eqnarray}
We see that the choice of fields is determined by the sign of $a$. Specifically, for $a > 0$, we use imaginary fields to address positive-definite cases, while for $a < 0$, we use real fields, which align with the negative definite cases discussed in the paper.
~~\\
~~\\

\subsection{2) Impact of Lattice Size on  $\alpha_c$}
\begin{figure}[t] \includegraphics[scale=0.45]{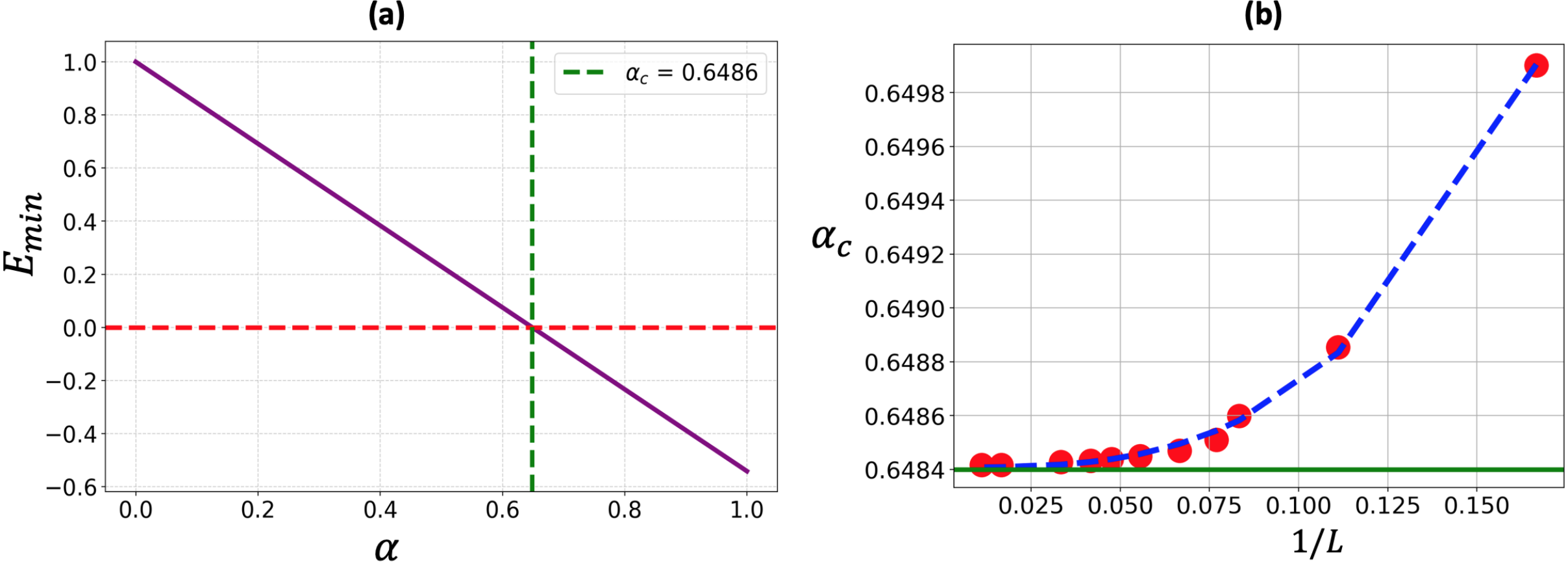} \caption{(a) The lowest eigenvalue, $E_{\text{min}}$, of the matrix $V$ as a function of $\alpha$ for a lattice with dimensions $L_1 = L_2 = L = 12$, with $U = 1$. 
(b) The critical value $\alpha_c$ as a function of lattice size, plotted for $L = 6, 9, 12, 15, 18, 21, 24, 27, 30, 60,$ and $90$. As lattice size increases, $\alpha_c$ converges to an asymptotic value around $0.6484$, shown by the green line.} \label{alpha_c_SM} \end{figure}

\indent Determining the exact value of $\alpha_c$ analytically is challenging, but numerical simulations offer strong evidence. To investigate, we numerically generated the matrix $V$ for various lattice sizes and analyzed the behavior of its lowest eigenvalue, $E_{\text{min}}$, as a function of $\alpha$, with $U$ set to 1 (since it serves only as a prefactor to $V$). At $\alpha = 0$, $V$ is diagonal, and all eigenvalues are positive, confirming it as positive-definite. As $\alpha$ increases, $E_{\text{min}}$ decreases and eventually crosses zero; beyond this crossing point, further increases in $\alpha$ lead to negative values for $E_{\text{min}}$, indicating that $V$ becomes indefinite.
This behavior is shown in panel (a) of Fig.~\ref{alpha_c_SM}, where we plot $E_{\text{min}}$ versus $\alpha$ for $L_1 = L_2 = L = 12$, finding that $\alpha_c = 0.6486$ in this case. We repeated this procedure for other lattice sizes, including $L = 6, 9, 12, 15, 18, 21, 24, 27, 30, 60$, and $90$, extracted the corresponding values of $\alpha_c$, and plotted them in panel (b) of Fig.~\ref{alpha_c_SM}. This plot shows that as lattice size increases, $\alpha_c$ approaches an asymptotic value close to $0.6484$, indicated by the green line.

~~\\
~~\\

\subsection{3) Establishing the Lower Bound of $\epsilon$ in the DD Method}
\indent Regarding the definitions of $V^-$ and $V^+$ in the DD method, note that we can rewrite $V$ by adding and subtracting an arbitrary matrix, chosen here as $\epsilon I$, where $I$ is the identity matrix. This gives $V = V + \epsilon I - \epsilon I$, allowing us to define $V^+ = V + \epsilon I$ and $V^- = -\epsilon I$. Clearly, for any $\epsilon > 0$, $V^-$ is negative definite. However, ensuring that $V^+$ is positive-definite is less straightforward.

\indent To illustrate this, consider two limits (assuming that $V$ is indefinite): when $\epsilon = 0$, we have $V^+ = V$, which is not positive-definite. In the opposite limit, $\epsilon \to \infty$, $V^+ \sim \epsilon I$, which is positive-definite. Thus, we expect that at some finite $\epsilon$ between these limits, $V^+$ becomes positive-definite. To identify this lower bound more precisely, we move to the diagonal representation of $V$.
As in the ED method, let $W$ be an orthonormal matrix that diagonalizes $V$, i.e., $V_d = W^T V W$: 
\begin{equation}
V_d = \text{Diag}(\lambda_1, \lambda_2, \dots, \lambda_N) \end{equation}
where $\lambda_1 < \lambda_2 < \dots < \lambda_N$ are the eigenvalues of $V$, with $\lambda_1$ as the most negative eigenvalue. Since $W^T I W = I$, we obtain $V^{+}$ in the new basis as $V_d^{+}$: 
\begin{equation} 
V^{+}_d = \text{Diag}(\lambda_1+\epsilon, \lambda_2+\epsilon, \dots, \lambda_N+\epsilon). \end{equation} 
To make all eigenvalues of $V^{+}_{d}$ positive-definite, we require $\epsilon > -\lambda_1$. Thus, we conclude that $\epsilon_{\text{min}} = -\lambda_1$ sets the lower bound on $\epsilon$. Finally, we can transform this back to the original basis using $V^{+} = W V_d^{+} W^T$.

~~\\
~~\\
\section{4) Projective Auxiliary-Field Quantum Monte Carlo Method}

\indent The Projective Auxiliary-Field Quantum Monte Carlo (AFQMC) method is a zero-temperature algorithm designed to access the ground state properties of quantum many-body systems. This approach projects the ground state wave function $| \Psi_0 \rangle$ from an initial pair of trial wave functions, $| \Psi_{T; L} \rangle$ and $| \Psi_{T; R} \rangle$, which are chosen to have a nonzero overlap with $| \Psi_0 \rangle$: 
\begin{eqnarray}
 \langle \Psi_{T; L} | \Psi_0 \rangle \neq 0 \quad \text{and} \quad \langle \Psi_{T; R} | \Psi_0 \rangle \neq 0. 
\end{eqnarray}

\indent To obtain the ground state expectation value of any observable $\hat{O}$, we propagate the trial wave functions along the imaginary time axis, as follows: 
\begin{eqnarray} 
\langle \Psi_0 | \hat{O} | \Psi_0 \rangle = \lim_{\Theta \to \infty} \frac{\langle \Psi_{T; L} | e^{-\Theta \hat{H}} e^{-(\beta-\tau) \hat{H} }\hat{O} e^{-\tau \hat{H}}e^{-\Theta \hat{H}} | \Psi_{T; R} \rangle}{\langle \Psi_{T; L} | e^{-(2\Theta+\beta) \hat{H}} | \Psi_{T; R} \rangle}, \end{eqnarray}
where $\Theta$ defines the imaginary time evolution length used to project out the ground state. In the algorithm, a large but finite $\Theta$ is selected to ensure convergence to the ground state within statistical uncertainty. Observables can be calculated at different imaginary time $\tau$ ranges from $0$ to $\beta$, allowing for both equal-time and time-displaced measurements.
This projection process involves the Trotter-Suzuki decomposition of the Hamiltonian $\hat{H}$, separating the exponential of the kinetic and interaction terms. The interaction term is then decoupled using a Hubbard-Stratonovich transformation, which introduces a set of auxiliary fields. Monte Carlo sampling is performed over these fields, enabling efficient evaluation of the projected ground state expectation values.

\indent In our simulations, we use the non-interacting ground state of $\hat{H}_T$ as the trial wave functions $| \Psi{T; L} \rangle$ and $| \Psi_{T; R} \rangle$, as this choice reliably overlaps with the ground state for the models considered. This method provides a robust approach to calculate ground-state properties accurately, as long as convergence in $\Theta$ is ensured.

~~\\
~~\\

\subsection{5) Behavior of Average Sign in Eigenvalue-Based Decomposition (ED) Method}

\begin{figure}[t] \includegraphics[scale=0.4]{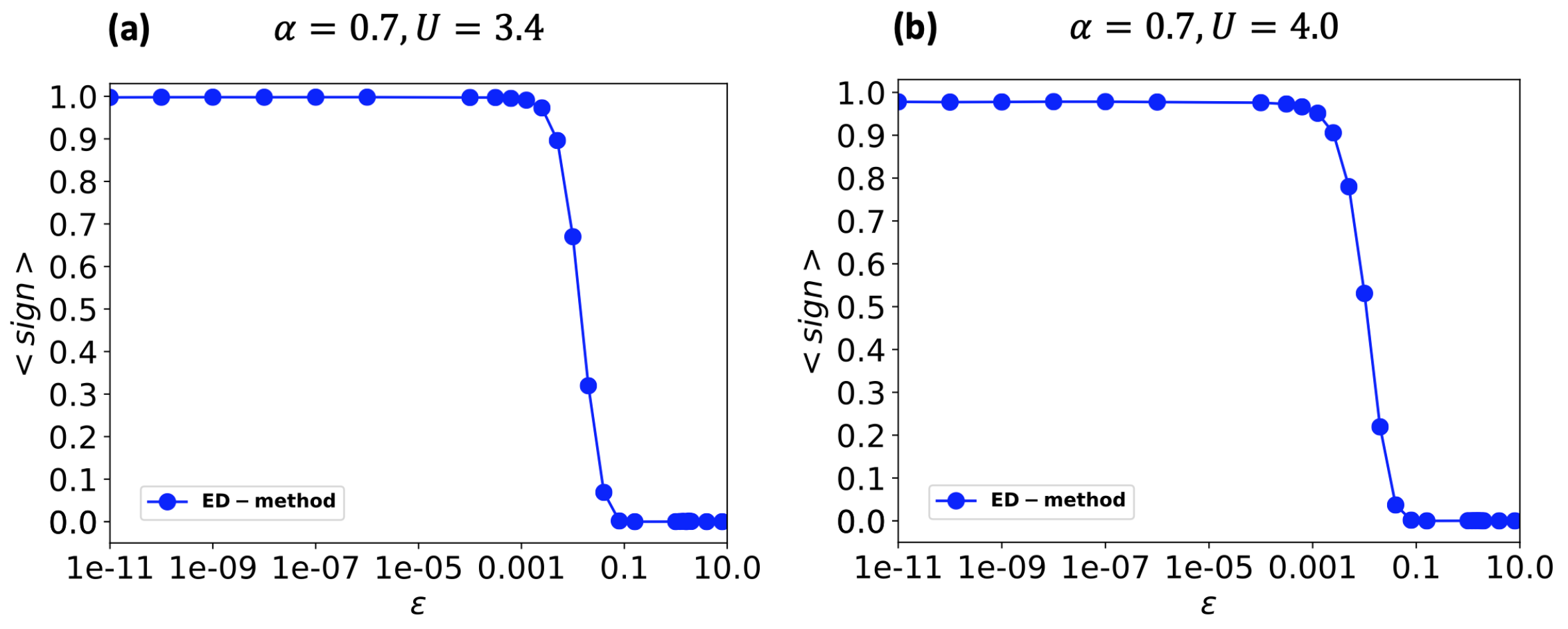} \caption{The average sign for the eigenvalue-based decomposition of the interaction matrix for a lattice with $\alpha=0.7$, $L=9$, $\beta=12.0$, and (a) $U=3.4$, (b) $U=4$.} \label{sign_SM} \end{figure}

\indent In the main text, we presented results for ⟨sign⟩ using the parameter set $\alpha = 0.8$, and $L = 9$, $\beta = 12$, and $U = 3.4$. Here, we consider two different sets: $\alpha = 0.7$, $L = 9$, $\beta = 12$, $U = 3.4$ and $U = 4.0$ to demonstrate similar behavior, as shown in Fig.~\ref{sign_SM} (comparable results were found across other parameter sets).
As shown, ⟨sign⟩ improves as $\epsilon$ decreases and converges below a certain threshold.
It is worth noting that we expect with a different choice of the matrix $V$, the point at which ⟨sign⟩ convergence begins may differ.

~~\\
~~\\
\subsection{6) Impact of $\epsilon$ on SM-CDW Phase Boundary and Average Sign Drop}

\begin{figure}
\includegraphics[scale=0.75]{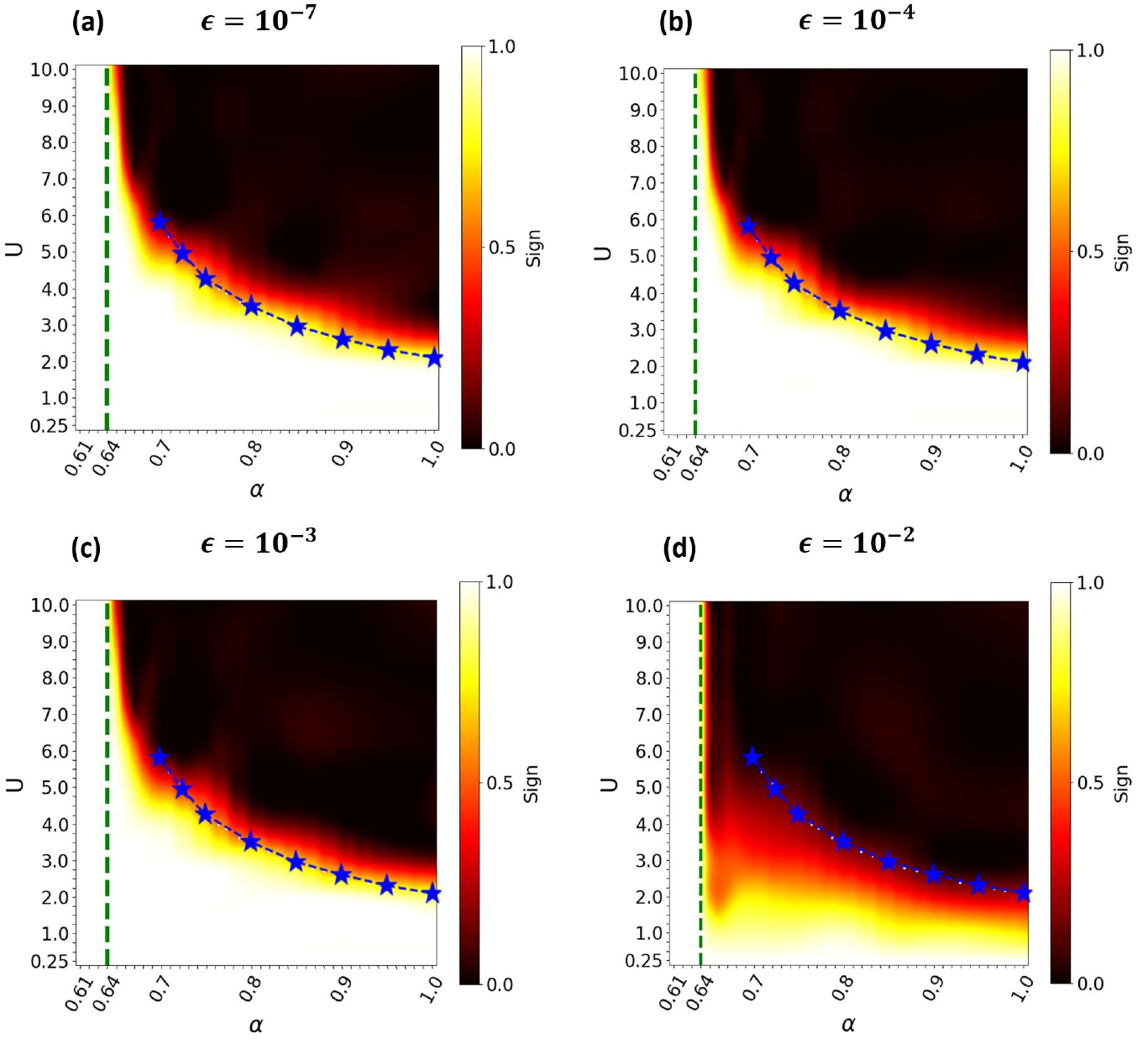}
\caption{The contour plot illustrates the behavior of the average sign, ⟨sign⟩, for $\beta=12$, $L=12$ as a function of $U$ and $\alpha$ across various values of $\epsilon$. Within the convergence range of ⟨sign⟩ with respect to $\epsilon$, we observe that the alignment between the SM-CDW phase boundary and the sharp drop in ⟨sign⟩ remains stable and consistent.}
\label{heatmap_e_SM}
\end{figure}

\indent To better illustrate how the choice of $\epsilon$ may affect the contour plot, we present additional data for the same parameters used in the main text but with $\epsilon = 10^{-7}, 10^{-4}, 10^{-3},$ and $10^{-2}$, as shown in Fig.~\ref{heatmap_e_SM}. As noted, within the convergence range of ⟨sign⟩, the SM-CDW phase boundary consistently aligns with the sharp drop in ⟨sign⟩, showing minimal dependence on $\epsilon$. However, for $\epsilon = 10^{-2}$, which is near the convergence threshold (see Fig.~\ref{sign_SM}), this alignment does not hold. Therefore, we conclude that the correspondence between the SM-CDW phase boundary and ⟨sign⟩ is robust over a broad range of $\epsilon$ values.
Please note that the contour plot for $\epsilon = 10^{-10}$ presented in the main text was generated using a higher density of data points. Any minor discrepancies between the plots here and those in the main text are due to this difference. 

~~\\
~~\\

\subsection{7) Impact of Lattice Size and Inverse Temperature on SM-CDW Phase Boundary and Average Sign Drop}

\indent In the main text, we presented results for $L = 12$ and $\beta = 12$ in Fig.~\ref{heatmap}. Here, we extend this analysis by generating contour plots for $L = 9$, $\beta = 12$ and $L = 12$, $\beta = 6$, to examine the effects of lattice size and inverse temperature, respectively. These plots reveal some minor deviations, demonstrating that the alignment of the SM-CDW phase boundary with the sharp drop in ⟨sign⟩ remains consistent across these variations.
\begin{figure}
\includegraphics[scale=0.52]{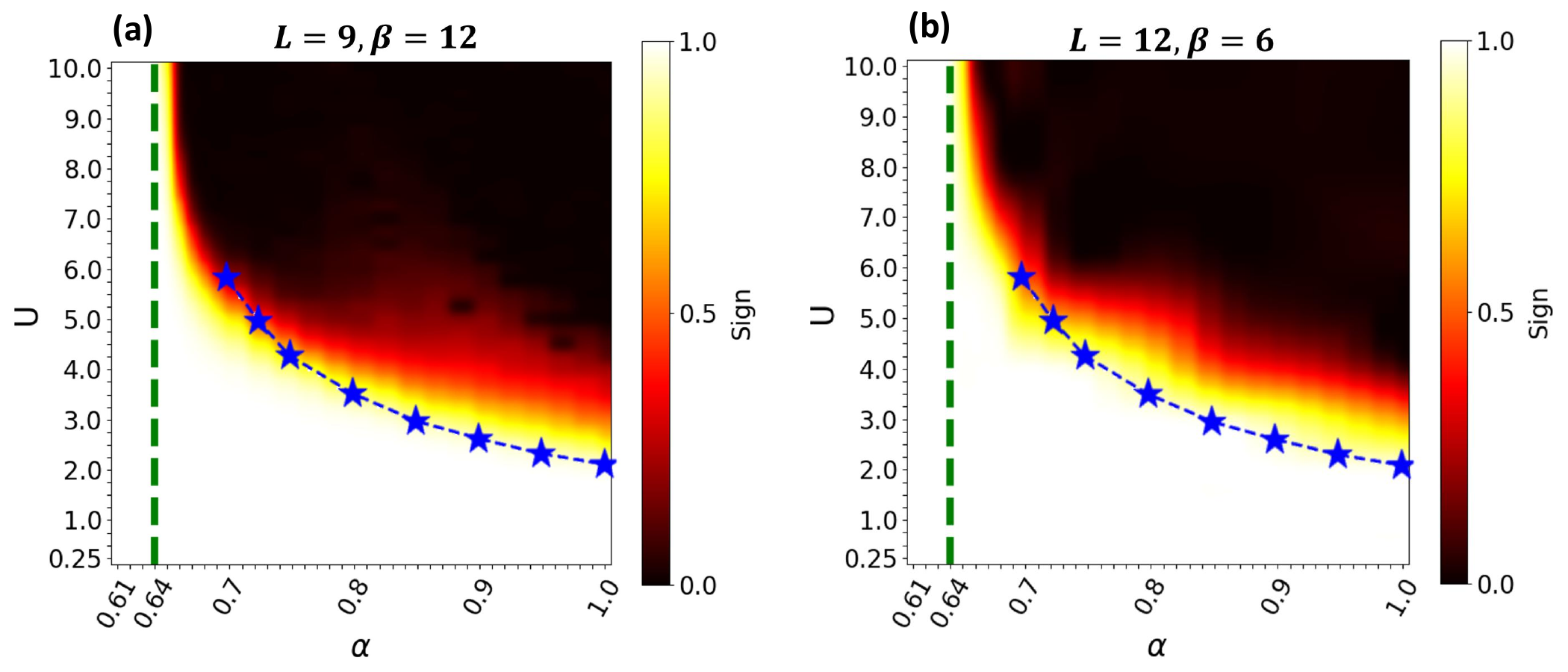}
\caption{The contour plot illustrates the behavior of the average sign, ⟨sign⟩, across various values of $U$ and $\alpha$. We observe that the alignment between the SM-CDW phase boundary and the sharp drop in ⟨sign⟩ remains stable and consistent.}
\label{heatmap_Lb_SM}
\end{figure}
~~\\
~~\\
\subsection{8) Application of Eigenvalue Decomposition in the Square Lattice LRC Model: Managing Indefinite Interactions}

\indent As discussed, our eigenvalue decomposition (ED) approach for handling indefinite interaction matrices is not limited to the honeycomb lattice. Specifically, in the long-range Coulomb (LRC) model on a square lattice, the interaction matrix becomes indefinite for $\alpha > 0.61$. While screening effects typically reduce the impact of long-range interactions, often making it sufficient to consider only nearest-neighbor interactions (NNI) for phase diagram analysis, we find it valuable to test our method on this lattice for completeness.

\indent Previous studies indicate that when local interactions dominate, the spin-density wave (SDW) phase is favored, whereas charge-density wave (CDW) phases emerge when NNI dominates. Additionally, the semi-metal (SM) phase is almost entirely suppressed at finite values of both local and non-local interactions. Therefore, even with LRC interactions, we anticipate the onset of the SDW phase at very low values of $U$ for $\alpha < 0.61$, and the CDW phase at similarly low $U$ values for $\alpha > 0.61$. However, as previously mentioned, the Auxiliary-Field Quantum Monte Carlo (AFQMC) method fails to handle this regime without our ED approach. 
It is worth noting that our preliminary analysis, which will be published soon, reveals considering the LRC interaction, enhances the SM phase at small values of $U$. 

\begin{figure}[t]
\includegraphics[scale=0.45]{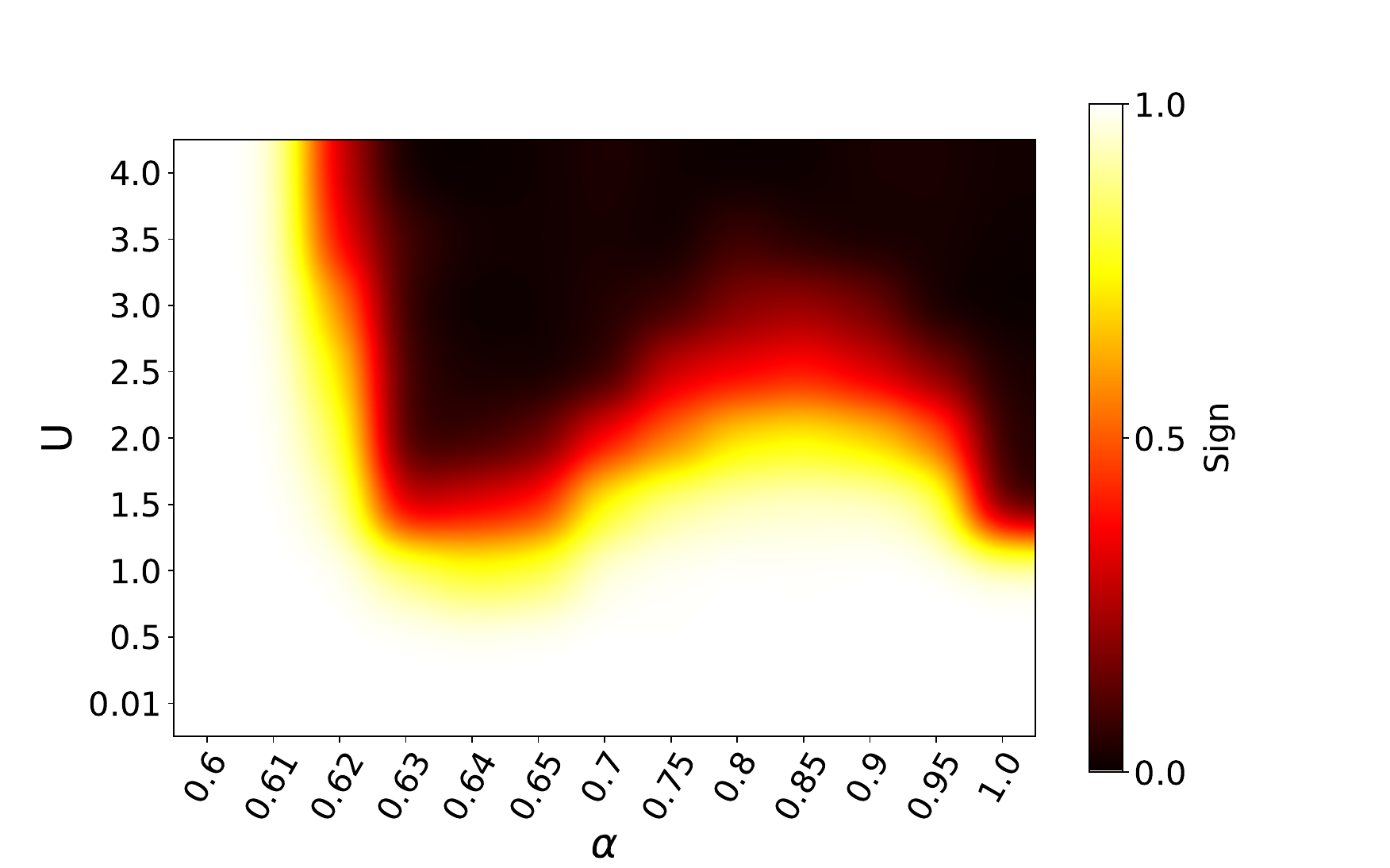}
\caption{The heat map illustrates the sign behavior for a square lattice with $L_x = L_y =20$ at $2\Theta + \beta = 45$, showing the dependence on the local interaction parameter $U$ and the long-range Coulomb interaction parameter $\alpha$.}
\label{heatmap_SL}
\end{figure}

\begin{figure}[t]
\includegraphics[scale=0.42]{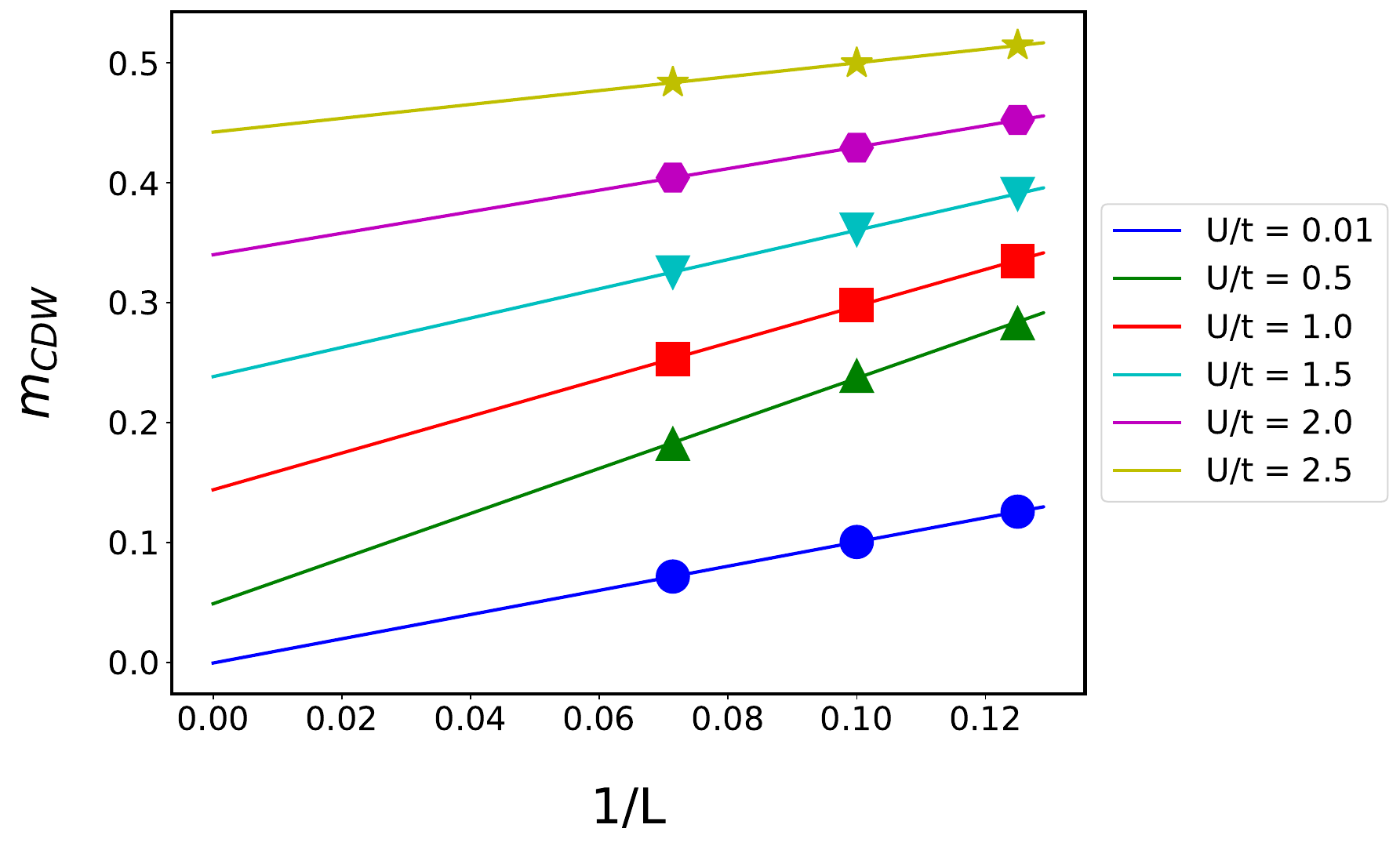}
\caption{CDW order parameter versus the inverse of linear length for $\alpha = 0.85$ at $2\Theta + \beta = 40$. Even for small values of $U$ a nonzero order parameter, signaling the CDW phase, is evident.}
\label{a_85_SL}
\end{figure}

\indent To assess the effectiveness of our method, we have plotted the behavior of the average sign in Fig.~\ref{heatmap_SL} for a square lattice with $L_x = L_y = L = 20$ and $2\Theta + \beta = 45$, using Projective AFQMC.
Remarkably, the sign problem remains manageable even in these extreme cases. As explained earlier, identifying the SM and CDW phases requires only a range where the sign is close to 1, i.e., $U<1$. This implies that severe sign issues at higher $U$ values are not a concern in this context, allowing us to reliably gather the necessary data.
For instance, in Fig.~\ref{a_85_SL}, we have plotted the $m_{\text{CDW}}$ for $\alpha = 0.85$ at $2\Theta + \beta = 40$. 
Evidently, a nonzero $m_{\text{CDW}}$ is observed even for $U = 0.5$.

\end{document}